\documentclass[aps,prl,twocolumn,superscriptaddress,reprint]{revtex4-1}

\usepackage[ansinew]{inputenc}
\usepackage[T1]{fontenc}
\usepackage{ae,aecompl}
\usepackage[english]{babel}
\usepackage{graphicx,pict2e}
\usepackage{hyperref,amsmath}
\usepackage{epstopdf}
\usepackage{color}
\usepackage{multirow}
\usepackage[per-mode=reciprocal,range-units = single,range-phrase=\ -\ ,exponent-product = \cdot]{siunitx}
\usepackage[normalem,normalbf]{ulem}

\begin{document}

\author{M. Naumann}
\author{F. Arnold}
\author{Z. Medvecka}
\author{S.-C. Wu}
\author{V. S\"uss}
\author{M. Schmidt}
\affiliation{Max Planck Institute for Chemical Physics of Solids, 01187 Dresden, Germany} 
\author{B. Yan}
\affiliation{Department of Condensed Matter Physics, Weizmann Institute of Science, Rehovot, Israel}
\author{N. Huber}
\affiliation{Physik-Department, Technische Universit\"at M\"unchen, 85748 Garching, Germany}
\author{L. Worch}
\affiliation{Physik-Department, Technische Universit\"at M\"unchen, 85748 Garching, Germany} 
\author{M.~A. Wilde}
\affiliation{Physik-Department, Technische Universit\"at M\"unchen, 85748 Garching, Germany} 
\author{C. Felser}
\author{Y. Sun}
\affiliation{Max Planck Institute for Chemical Physics of Solids, 01187 Dresden, Germany} 
\author{E. Hassinger}
\affiliation{Max Planck Institute for Chemical Physics of Solids, 01187 Dresden, Germany}
\affiliation{Physik-Department, Technische Universit\"at M\"unchen, 85748 Garching, Germany}

\title{Weyl nodes close to the Fermi energy in NbAs}

\date{\today}

\begin{abstract}
The noncentrosymmetric transition metal monopnictides NbP, TaP, NbAs and TaAs are a family of Weyl semimetals in which pairs of protected linear crossings of spin-resolved bands occur. These so-called Weyl nodes are characterized by integer topological charges of opposite sign associated with singular points of Berry curvature in momentum space.
In such a system anomalous magnetoelectric responses are predicted, which should only occur if the crossing points are close to the Fermi level and enclosed by Fermi surface pockets penetrated by an integer flux of Berry curvature, dubbed Weyl pockets. 
TaAs was shown to possess Weyl pockets whereas TaP and NbP have trivial pockets enclosing zero net flux of Berry curvature. Here, via measurements of the magnetic torque, resistivity and magnetisation, we present a comprehensive quantum oscillation study of NbAs, the last member of this family where the precise shape and nature of the Fermi surface pockets is still unknown. We detect six distinct frequency branches, two of which have not been observed before. A comparison to density functional theory calculations suggests that the two largest pockets are topologically trivial, whereas the low frequencies might stem from tiny Weyl pockets. The enclosed Weyl nodes are within a few meV of the Fermi energy.
\end{abstract}

\maketitle
\section{Introduction}
\begin{figure}[t]
\centering
\includegraphics[width=0.8\columnwidth]{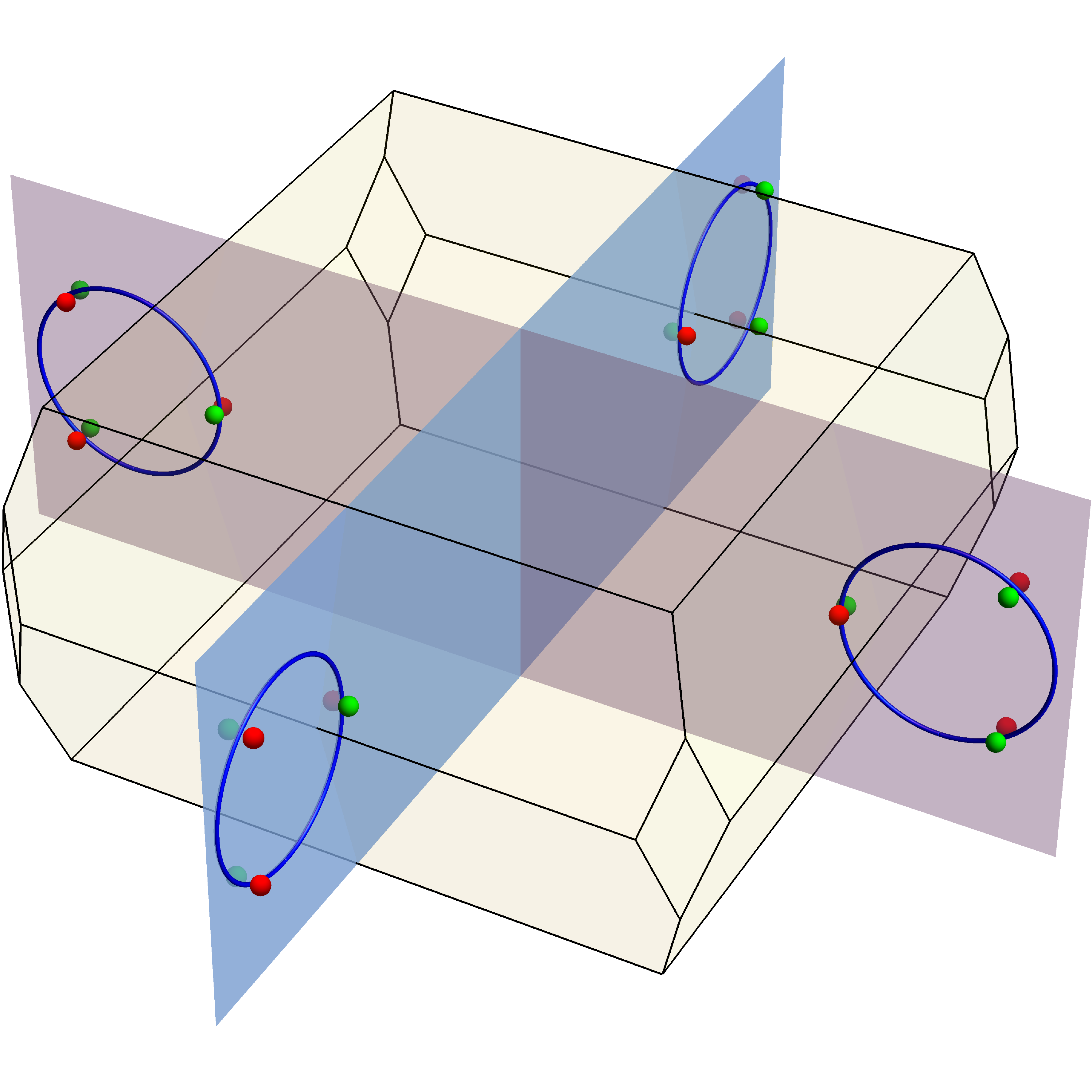}
\caption{{\bf Sketch of nodal rings and Weyl point positions in momentum space.} Nodal rings (blue) in the absence of SOC. Upon inclusion of SOC the nodal rings are gapped out and give rise to the Weyl points with different chirality (red and green) located off the mirror planes. The BZ contains four nodal rings with three pairs of Weyl points each.}
\label{fig:k-space-sketch}
\end{figure}

In the past years extensive studies have been performed to investigate and unveil the chiral massless Weyl fermions in the transition metal monopnictides MX (M=Ta, Nb) (X=P,As) and other proposed Weyl semimetal (WS) candidates\cite{yan_topological_2017,armitage_weyl_2018}. In the isostructural MX compounds, four nodal rings related by $C_4$ symmetry are predicted in the first Brillouin zone (BZ)(blue rings in \textbf{Figure~\ref{fig:k-space-sketch})}. Upon inclusion of spin-orbit-coupling (SOC), the nodal rings gap out and only three chiral pairs of Weyl points off the mirror planes remain gapless (red and green points in Figure~\ref{fig:k-space-sketch}). Early observations of the linear band-crossings in the bulk and topological Fermi arc surface states by angle-resolved photoemission spectroscopy were able to confirm the existence of Weyl fermions in all four materials of this family \cite{wan_topological_2011,lv_experimental_2015,xu_discovery_2015,yang_weyl_2015}, setting the stage for investigations of their relevance for the materials' electronic properties. Magnetotransport experiments tried to uncover the Adler-Bell-Jackiw anomaly, i.e., a negative longitudinal magnetoresistance when the electric and magnetic field are applied in parallel  \cite{adler_axial-vector_1969,bell_pcac_1969}. However, initial findings of negative magnetoresistance \cite{huang_weyl_2015,zhang_signatures_2016,yang_weyl_2015,du_large_2016,wang_helicity-protected_2016} later turned out to be caused by a phenomenon known as current jetting, which is observed in materials with high charge carrier mobility \cite{reis_search_2016,arnold_negative_2016}. The intrinsic longitudinal magnetoresistance is found to be dominated by classical orbital effects and weak antilocalisation \cite{Naumann_2020}, and the absence of the Adler-Bell-Jackiw magnetotransport might be explained by the distance of the Weyl nodes to the Fermi energy \cite{johansson_chiral_2019}.

In all members of this family, small Fermi surface pockets evolve along the nodal ring and quantum oscillation experiments are able to determine the precise shape and nature of the pockets. Depending on the distance of the Weyl nodes to the Fermi energy in the given electronic structure, either pockets enclosing zero total Berry flux or integer non-zero Berry flux occur. The latter situation occurs when separate Fermi surface pockets enclose each Weyl node of a pair, in which case we name them Weyl pockets. Combined quantum oscillation and density functional theory studies were able to indirectly prove the existence of Weyl pockets in TaAs \cite{arnold_chiral_2016}, whereas their existence was dismissed in TaP \cite{arnold_negative_2016} and NbP \cite{klotz_quantum_2016}. NbAs, however, is the last compound in this series where the Fermi surface is not completely established and the existence of Weyl pockets is under debate. Via magnetoresistance measurements, four quantum oscillation frequency branches have been detected and ascribed to $C_4$-related pairs of either topologically trivial electron and a non-trivial hole pocket \cite{luo_electron-hole_2015}, or to trivial nested electron and hole pockets \cite{Komada_2020}. These two types of larger pockets were confirmed by magnetic torque measurements and a kink of the magnetization at the quantum limit was ascribed to the zero energy Landau level characteristic for topological materials \cite{moll_magnetic_2016}.

Here, we present an experimental study of quantum oscillations in NbAs. Via magnetoresistance as well as magnetization and magnetic torque experiments in combination with DFT calculations, we are able to draw a more detailed picture of its Fermi surface. The main findings are that it consists of at least one further type of small pocket, additionally to the two known larger ones, with frequencies down to $F\approx\SI{0.7}{\tesla}$. The two large pockets are crescent-shaped topologically trivial electron and hole pockets whereas the small pocket represents a candidate for Weyl pockets with nodal points that are extremely close to the Fermi level.

\section{Experimental and computational methods}
Crystals of NbAs were grown by chemical vapor transport~(CVT) using iodine as transport agent. The polycrystalline material has been synthesised in a first step by a direct reaction of micro-crystalline powders of the elements niobium (Alfa Aesar \SI{99.99}{\percent}) and arsenic (Alfa Aesar \SI{99.9999}{\percent}). Starting from this microcrystalline powder, NbAs crystallised in a temperature gradient from \SI{940}{\celsius}~(source) to \SI{1060}{\celsius}~(sink) and a transport agent concentration of \SI{13.5}{\milli\gram\per\cm\cubed} iodine (Alfa Aesar \SI{99.998}{\percent}). Crystals obtained by CVT were characterised by electron probe micro analysis and using X-ray diffraction techniques.
Laue spectroscopy was used to confirm the single crystalline structure and orientation.

Magnetisation measurements were performed on a Quantum Design SQUID vibrating sample magnetometer between $T = 2$\,K and 20\,K in magnetic fields up to $B=7$\,T. For each angle, the sample was glued in the required orientation with an angular accuracy of $\pm 5^\circ$.

Transport measurements were carried out in an Oxford Instruments dilution refrigerator at $T=\SI{0.035}{\kelvin}$ in fields up to \SI{15}{\tesla} using a rotation stage. Data acquisition via National Instruments PXI cards and digital Lock-In procedures were used in this setup. 

The magnetic torque $\tau= m \times B$ was measured using CuBe foil cantilevers with a capacitive readout similar to the design in Refs.~\cite{Wilde2008,Wilde2010} in an Oxford Instruments $^3$He-insert at $T=\SI{0.25}{\kelvin}$ in magnetic fields up to \SI{15}{\tesla} and using TqMag piezoelectric torque cantilevers \cite{Rossel_1996} in a PPMS down to \SI{2}{\kelvin} and in magnetic fields up to 14\,T.

The magnetic field was rotated in the $(001)$, $(100)$ and $(110)$ planes. The field angles are defined with respect to the high-symmetry axis as indicated in \textbf{Figure~\ref{fig:angdepfreq}}c.

The electronic structure of NbAs was calculated using the Vienna ab-initio simulation package \cite{Kresse_1996}. NbAs crystallises in the space group I$4_1$md (\#109). The lattice parameters used were $a=b=3.4517$\,\AA, $c=11.680$\,\AA\,\,. The 4a Wyckoff positions of both Nb and As were allowed to relax in the calculations leading to (0,0,0) for Nb and (0,0,0.417) for As, respectively. The exchange correlation energy was considered on the level of the generalized gradient approximation (GGA), following the Perdew-Burke-Ernzerhof parametrization scheme \cite{Perdew96}. The energy cutoff of the plane waves was set to 450 eV.
The resulting band structure is in general consistent with the results published before \cite{lee_fermi_2015}.  For accurate Fermi surfaces the bands were interpolated using maximally localized Wannier functions \cite{Mostofi_2008} in dense k-grids of $300 \times 300 \times 300$ points in the full BZ. 

\begin{figure}[t]
\centering
\includegraphics[trim= 20 20 0 0, width=.8\columnwidth]{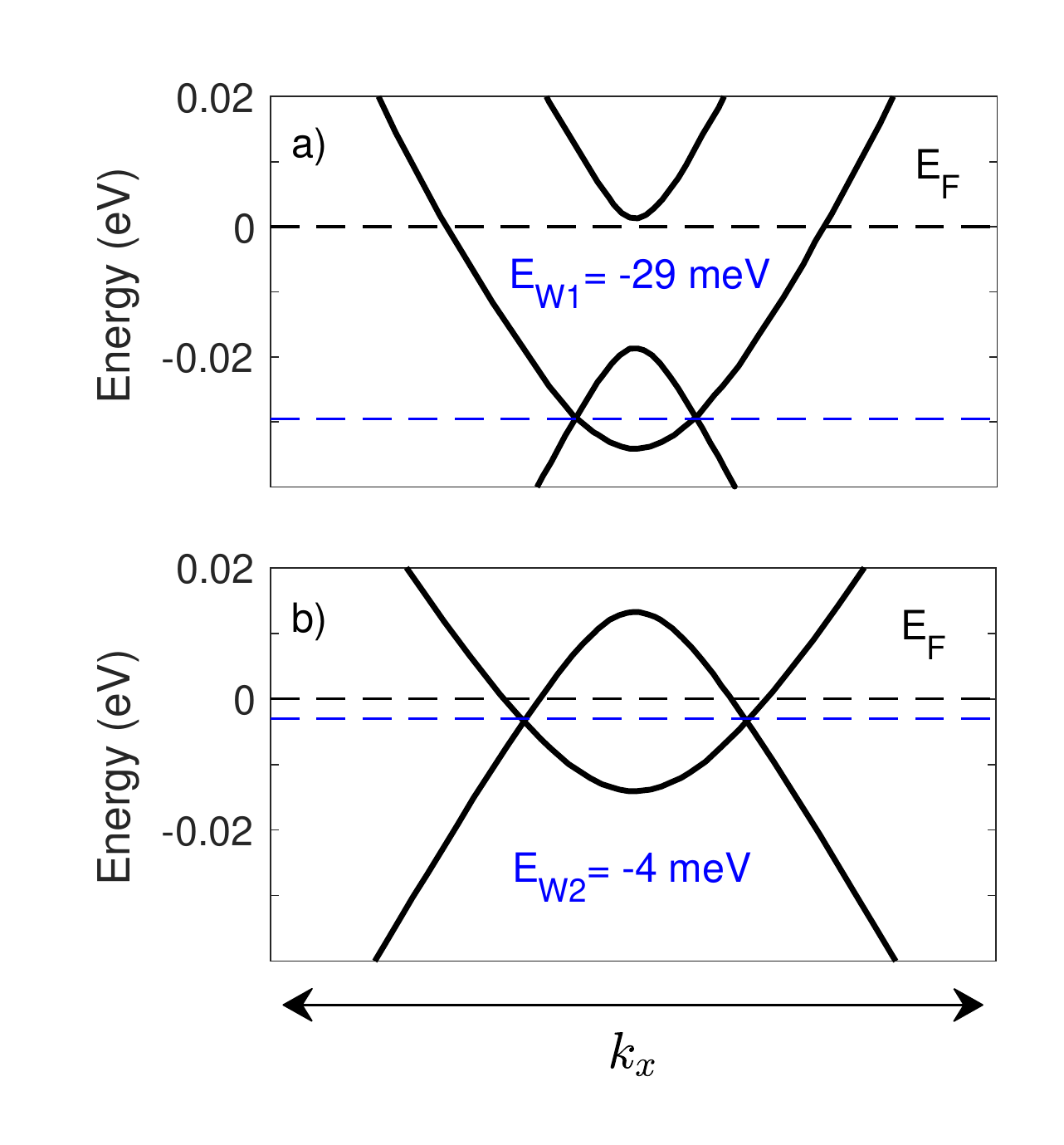}
\caption{{\bf Bandstructure of NbAs near the Weyl nodes.} The bandstructure of NbAs in the GGA with SOC along the lines connecting pairs of Weyl nodes (a) for W1 Weyl nodes  located at $(\pm0.0026,0.4817,0)$ in units of the conventional reciprocal lattice vectors and (b) for W2 Weyl nodes located at $(\pm0.0198,0.2811,0.5847)$ (and symmetry related) positions. The dashed lines indicate the Fermi energy (black) and the energy level of the Weyl node (blue), respectively.} 
\label{fig:bandstructure}
\end{figure}

Quantum oscillations appear, when electron energy levels are quantised in the presence of a magnetic field. With changing magnetic field, they pass through the Fermi energy and cause oscillations in the density of states which themselves lead to oscillations of resistivity or magnetisation. The oscillations are analysed using the standard Lifshitz-Kosevitch theory \cite{lifshitz_theory_1956,shoenberg_magnetic_1984}.  The quantum oscillation frequencies $F$ correspond to extremal Fermi surface cross sections $A$ perpendicular to the applied magnetic field direction via $F =( \hbar/2\pi e) \,A$. Thus, by rotating the magnetic field and comparing the angular dependence of the oscillation frequencies to those from bandstructure calculations, the Fermi surface topography can be obtained \cite{onsager_interpretation_1952,Ashby_2014}. The effective mass averaged over such extremal orbits is given by the temperature dependence of the oscillation amplitude $A\propto \chi/\sinh{(\chi)}$ where $\chi =  14.69\,m^\ast T/B$ \cite{shoenberg_magnetic_1984}. Here, $B$ is either a fixed field or the average reciprocal field of the analysed data window. In the second case, the field range was chosen to be as small as possible and the amplitude change with field was taken into account \cite{mercure_haas_2008}. For the quantum oscillation analysis, a background was subtracted from the measured data and a Fast Fourier transform (FFT) was performed on the residual oscillatory part.

\section{Results}

\subsection{Theoretical prediction}
In our calculations, the W1 Weyl points in the $k_{\text{z}}=0$-plane are located at $(\pm0.0026, 0.4817, 0)$ in units of the conventional reciprocal lattice vectors while the W2 Weyl points off the $k_{\text{z}}=0$-plane are at $(\pm0.0198, 0.2811, 0.5847)$ (and symmetry related) positions as shown in Figure \ref{fig:k-space-sketch} and \textbf{Figure \ref{fig:bandstructure}}. The W1 Weyl point pairs of opposite chirality reside \SI{29}{\milli\electronvolt} below the Fermi energy and are thus predicted to be enclosed by a single topologically trivial Fermi surface pocket (Figure \ref{fig:bandstructure}a). The W2 Weyl point pairs are predicted to be positioned only \SI{4}{\milli\electronvolt} below the Fermi level, giving rise to two separate chiral Weyl pockets (Figure \ref{fig:bandstructure}b and Figure \ref{fig:angdepfreq}a). We will discuss below that a similar scenario is indeed supported by our experimental results.

\subsection{Magnetisation along high-symmetry directions}

\begin{figure*}[t]
\centering
\includegraphics[trim=0 20 0 0 ,width=\textwidth]{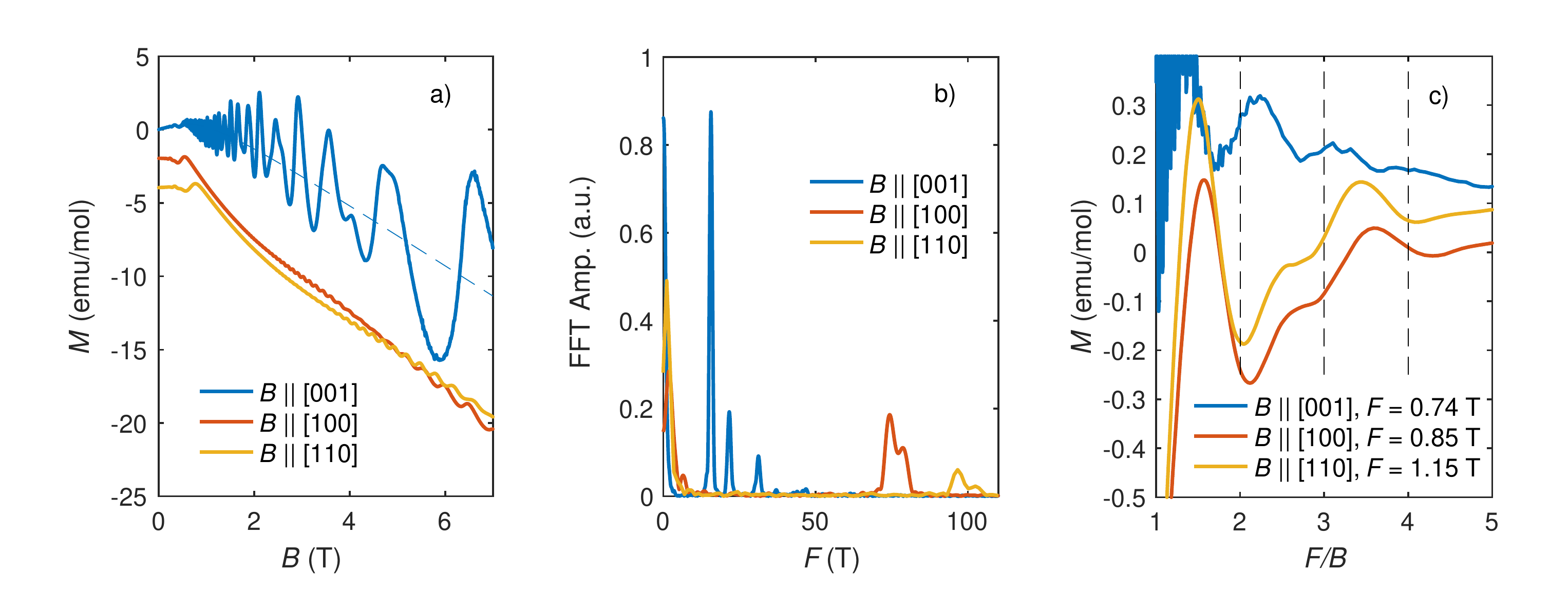}
\caption{{\bf Magnetisation of NbAs along the principal axes.} (a) Magnetisation at \SI{2}{\kelvin} for magnetic fields along the [001], [100] and [110] direction. Curves are offset for clarity. The thin dashed line is the fit acording to Eq. \ref{eq:boullion} of the magnetisation along the $c$ direction ([001]). (b) FFT of the data in (a) versus inverse magnetic field after subtracting the non-oscillatory background in the range 2\,T - 7\,T. (c) Magnetisation versus inverse field normalised by the small oscillation frequency as indicated.}
\label{fig:mag_data}
\end{figure*}

Let us start by looking at the magnetization at $T=2$\,K along the main crystallographic high-symmetry directions in \textbf{Figure~\ref{fig:mag_data}}(a). In all directions, strong quantum-oscillations are visible on top of a background signal consisting of two main contributions. The  strong diamagnetic contribution $M_\mathrm{dia}$, as expected for a semimetal, and a paramagnetic contribution $M_\mathrm{para}$ from a small amount of magnetic impurities. We model those contributions by using the following fit for the data for $B \parallel [001]$:
\begin{align}
      M &= M_\mathrm{para} + M_\mathrm{dia},\label{eq:boullion}\\
  M_\mathrm{para} &= N g J \mu_\mathrm{B} B_J(x),\ x=gJ\mu_\mathrm{B} B/k_\mathrm{B} T,\\
  B_J (x) &= \frac{2J+1}{2J}\coth\left ( \frac{(2J+1)x}{2J}\right ) -\frac{1}{2J}\coth \left ( \frac{x}{2J}\right ),\\
  M_\mathrm{dia} &= \chi B.
\end{align}
This yields a total magnetic moment of $J = \num{3.5}$, an impurity concentration $N = \SI{80}{ppm}$ and the diamagnetic susceptibility $\chi = -\SI{2.1}{emu\per\mol}$.

The strong quantum oscillations of high frequency are analysed by subtracting the background signal and taking the FFT of the oscillatory part in inverse fields (see Figure \ref{fig:mag_data}(b)). These frequency branches at $F = 15.8$\,T and $F = 21.8$\,T for $B\parallel [001]$ have been observed before and have been associated with trivial electron pockets and nontrivial hole pockets based on the effective mass and a phase shift \cite{luo_electron-hole_2015} or with trivial electron and hole pockets \cite{Komada_2020}. As we will show below, we come to the conclusion that both stem from trivial electron and hole pockets, as in \cite{Komada_2020}.

Additionally, our measurements reveal a smaller frequency in all directions that appears clearly at fields of $F \approx 1$\,T. A zoom on this field range is given in Figure~\ref{fig:mag_data}(c), where we normalised the inverse field range by the oscillation frequencies.

A small quantum oscillation frequency indicates a very small Fermi surface pocket, that will reach its quantum limit at a magnetic field of the order of the quantum oscillation frequency. Above this field, the Fermi surface pocket is depleted and the charge carriers redistribute to other pockets in a multiband system. In accordance with this, the oscillation with the small frequency stops after a final kink in the data.

\subsection{Temperature dependence and effective masses}

\begin{figure*}[t]
    \centering
    \includegraphics[trim=40 20 0 0 ,width=\textwidth]{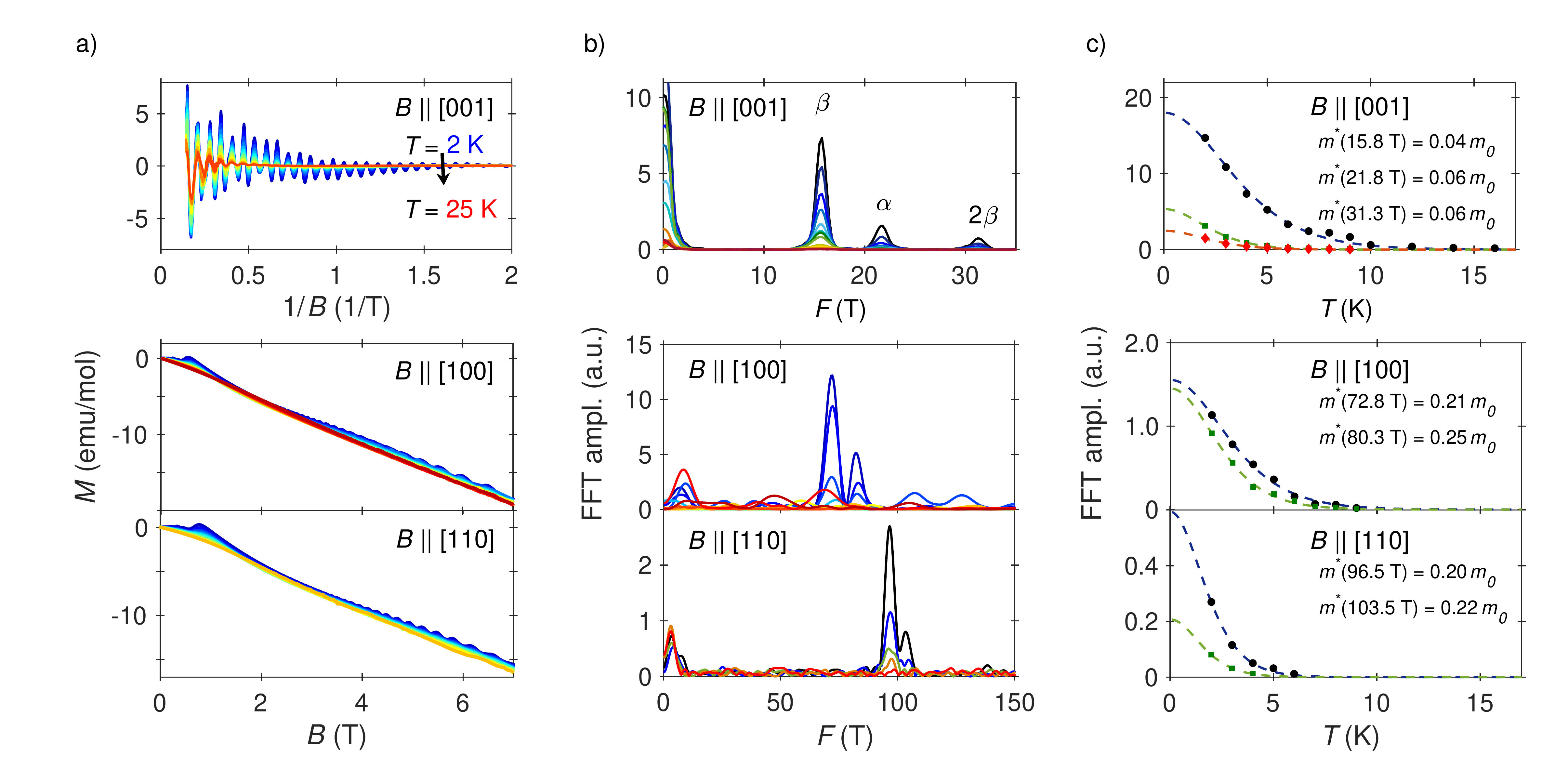}
    \caption{{\bf Temperature dependence of quantum oscillations.} (a) Magnetisation for different temperatures for the field directions as indicated. (b) FFT of the oscillatory part of the data in (a). (c) Temperature dependence of the oscillation amplitude obtained from the FFTs (dots) and the Lifshitz-Kosevich fit (dashed lines) with the resulting effective masses as indicated.}
    \label{fig:mass}
\end{figure*}

In a second step, we analyze the temperature dependence and determine the effective masses in \textbf{Figure \ref{fig:mass}}. The oscillatory part of the magnetisation as a function of inverse field for different temperatures is shown for $B\parallel [001]$ and the magnetisation at different temperature as a function of $B$ for $B\parallel [100]$ and $[110]$ in Figure \ref{fig:mass}(a). The FFTs of the oscillations are given in Figure \ref{fig:mass}(b). The resulting temperature dependence of the amplitudes and the Lifshitz-Kosevich fits with corresponding effective masses are depicted in Figure \ref{fig:mass}(c). The uncertainty of the effective masses is of the order of 10\,\%. The effective mass values are in agreement with those obtained before \cite{luo_electron-hole_2015} within the error bars. For the low frequencies, we were not able to determine the effective masses, because the signal is very small and the background changes strongly with temperature.

\subsection{Angular dependence}
\begin{figure*}[t]
\includegraphics[width=\textwidth]{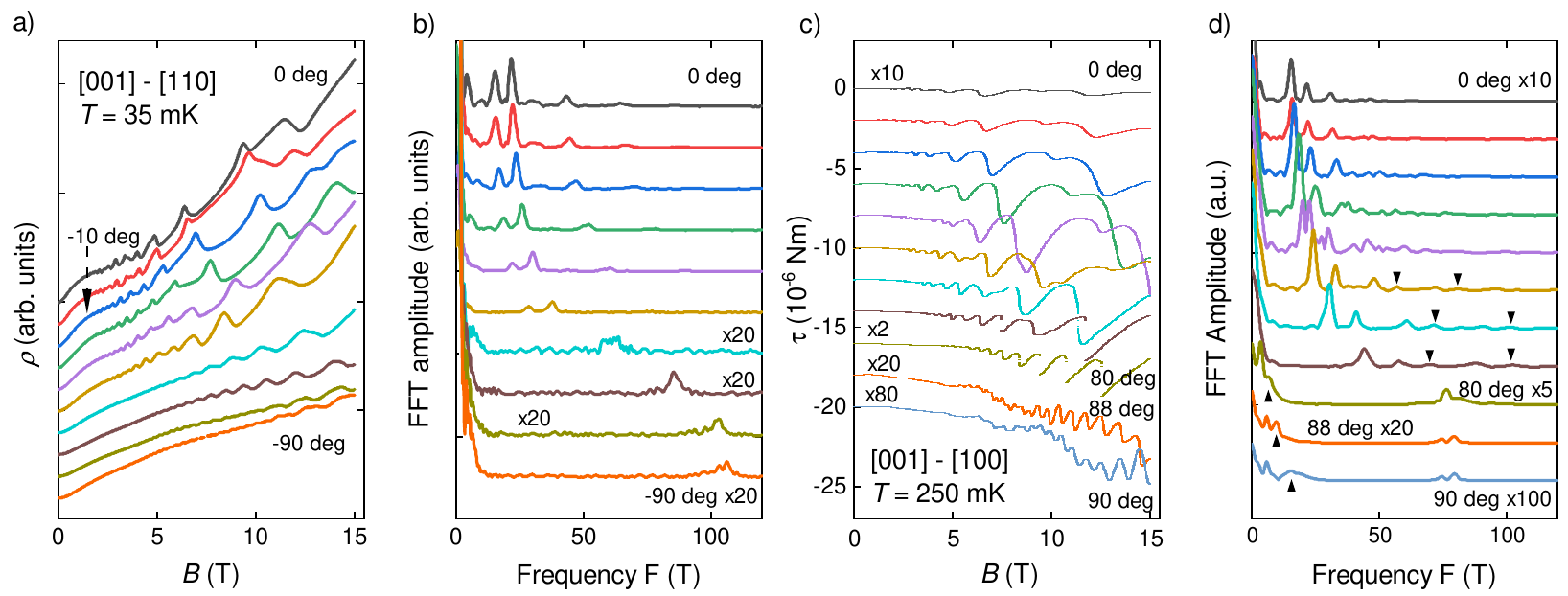}
\caption{{\bf Angle dependent quantum oscillation data.} All curves are offset for clarity and if not indicated otherwise, the angle was changed by 10 degrees between the curves. Scaling factors are applied as indicated. (a) Transverse resistivity at $T=\SI{35}{\milli\kelvin}$ for a rotation of the magnetic field in the (110) plane. (b) FFT of the data in (a), after subtraction of a polynomial background. The FFT range was \SIrange{1}{15}{\tesla}. (c) Magnetic torque at \SI{0.25}{\kelvin} in the (100) plane. (d) FFT of the data in (c). Downward triangles indicate the position of two additional high-frequency branches as observed before \cite{Komada_2020}. Upward triangles indicate the position of a new low-frequency branch. A moving average was removed as background and the FFT range was \SIrange{1}{15}{\tesla}.}
\label{fig:angdep}
\end{figure*}

\begin{figure}[t]
\includegraphics[width=.9\columnwidth]{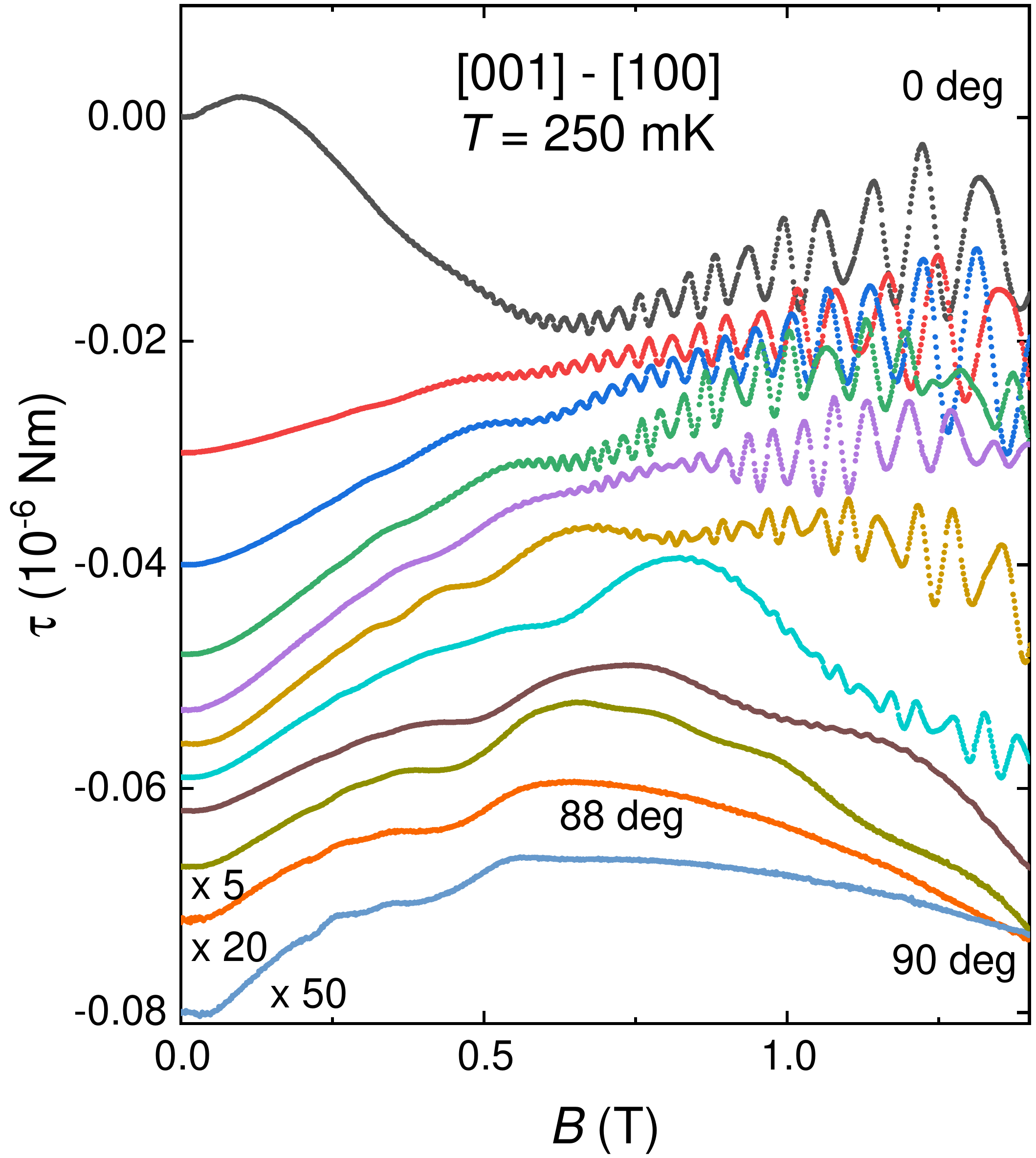}
\caption{{\bf Low-field angle dependent torque data.} Data are the same as in Figure \ref{fig:angdep}c. All curves are offset for clarity by different amounts and if not indicated otherwise, the angle was changed by 10 degrees between the curves. Scaling factors are applied as indicated.}
\label{fig:angdep2}
\end{figure}

\begin{figure}
\begin{center}
	\includegraphics[width=.7\columnwidth]{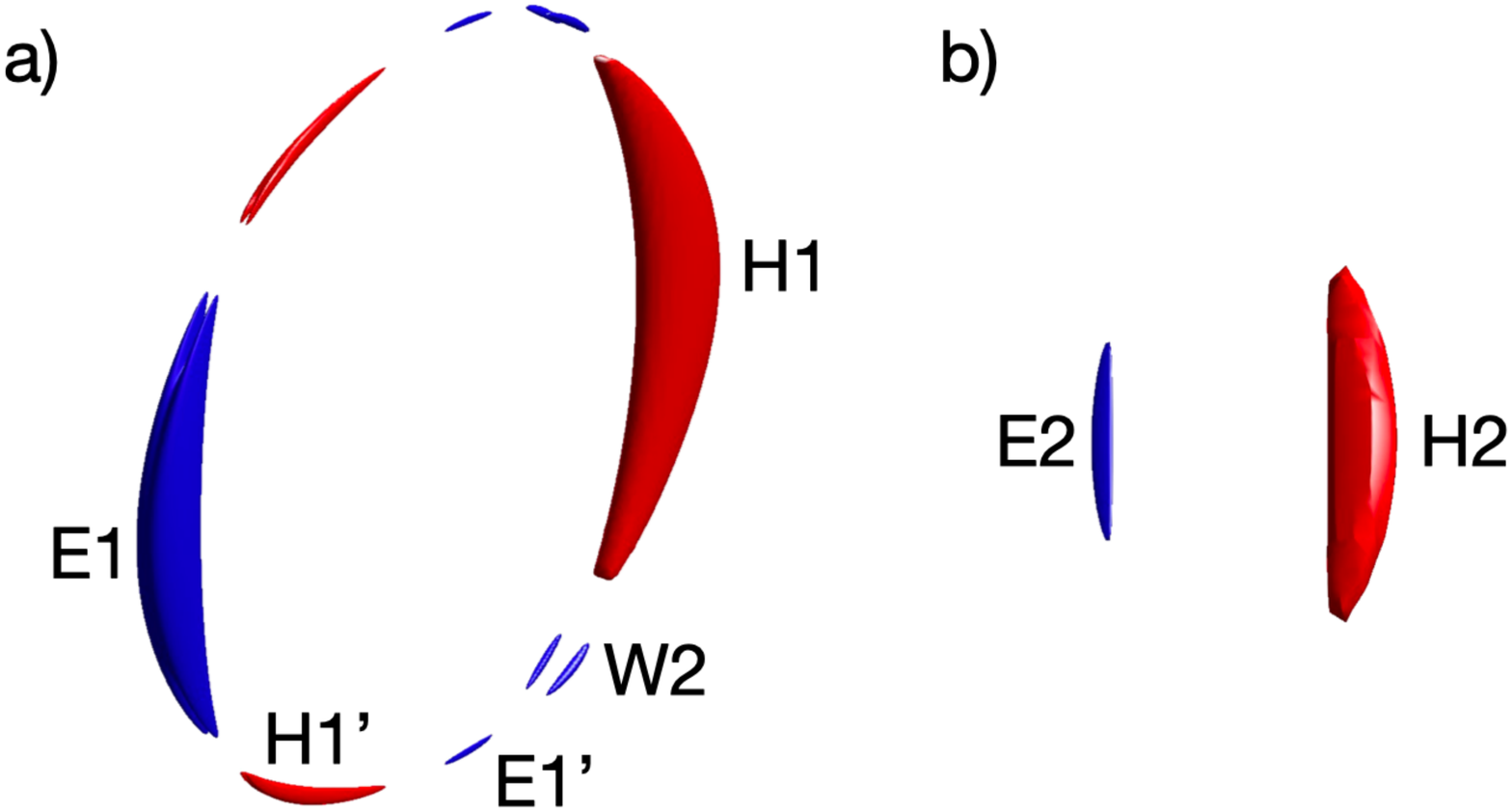}\\
	\includegraphics[width=.9\columnwidth]{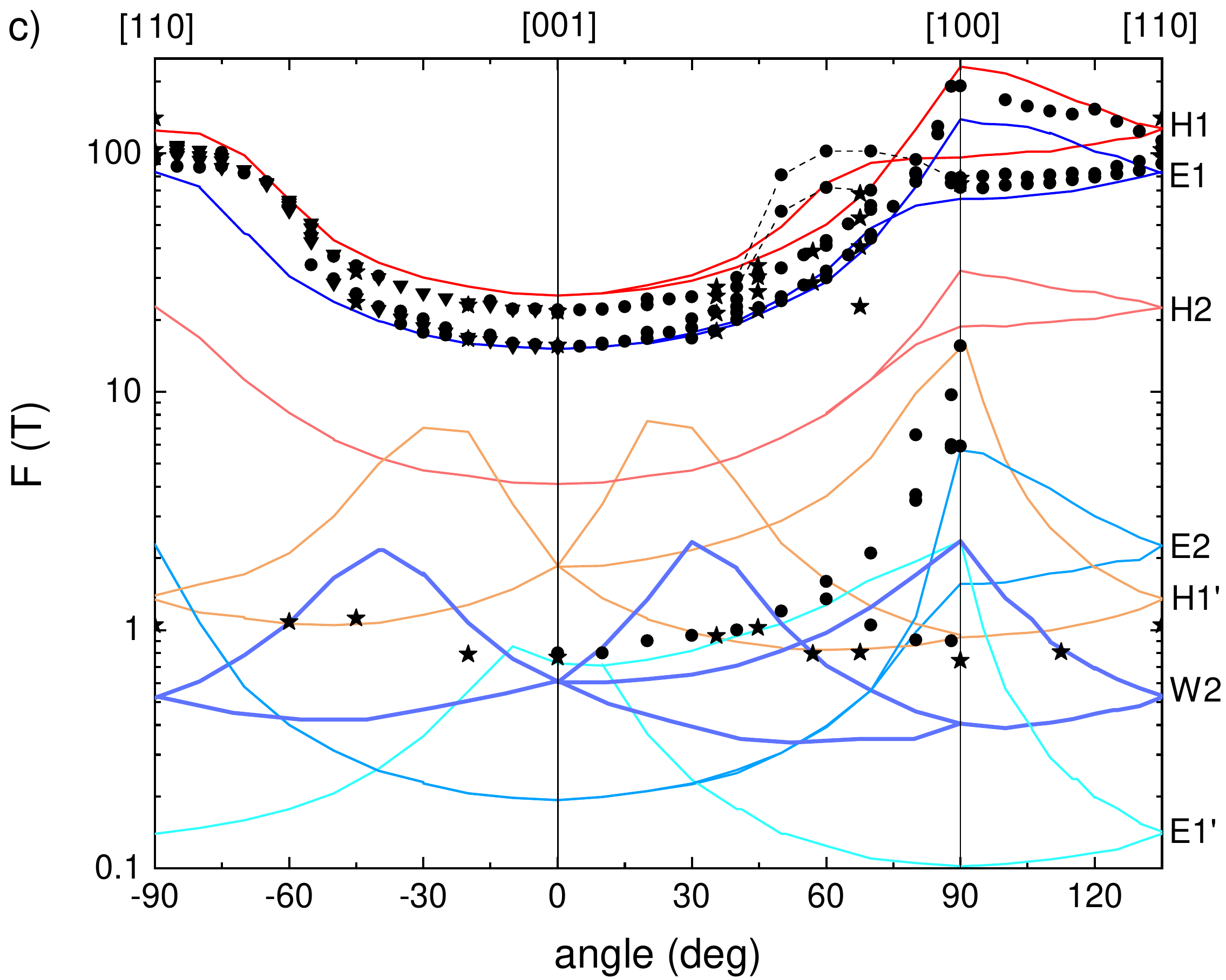}
\end{center}
\caption{{\bf Fermi surface and angular dependence of the quantum oscillation frequencies.} (a) Bird-eye's view on the Fermi surface pockets aligned along one nodal ring, where the E1 pocket is oriented towards the $\Gamma$ point. (b) Nested Fermi surface pockets in side-view: H2 is located inside H1 and E2 inside E1. (c) Angular dependence of quantum oscillation frequencies from torque (dots), resistivity (triangles) and magnetisation (stars) and from the DFT band structure (lines). Different full lines correspond to the Fermi surface pockets as indicated in (a) and (b) where reddish colors represent hole like surfaces H1, H2, H1' and blueish represent electron like Fermi pockets E1, E2, E1' and W2. The black dashed lines are a guide to the eye for the two splitted frequency branches.}
\label{fig:angdepfreq}
\end{figure}

\textbf{Figure \ref{fig:angdep}} shows the field dependence of resistivity and torque in NbAs taken at various angles for a rotation of the magnetic field as indicated. The resistivity in Figure \ref{fig:angdep}(a) exhibits metallic behaviour as reported before \cite{Luo_hard_2016}. The magnetoresistance in \SI{15}{\tesla} is around 100. This is large compared to normal metals but smaller than what was previously reported on NbAs pointing to a slightly lower crystal quality \cite{Luo_hard_2016}. Superimposed on the magnetoresistive background are pronounced quantum oscillations starting at around 0.8\,T. In the torque data shown in Figure \ref{fig:angdep}c and \textbf{Figure \ref{fig:angdep2}}, quantum oscillations are observable down to fields as low as 0.2\,T.
The FFT spectra of the resistivity and torque measurements at the accordant angles are shown in Figure \ref{fig:angdep}b and d. The frequencies found here for $B \parallel [001]$ agree with those observed in the magnetisation ($F=15.5$\,T and $F=21.5$\,T). Higher harmonics are also visible. For example, the third highest observed peak corresponds to the second harmonic of the 21.5\,T component for the resistivity and the \SI{15.5}{\tesla} component for the torque, each corresponding to the oscillation with largest amplitude. 
With increasing angle we observe a shift of the two main FFT peaks to higher frequencies and a characteristic splitting within the (001) tilting plane (downward triangles in Figure \ref{fig:angdep}d and dashed lines in Figure \ref{fig:angdepfreq}c). This behavior is in agreement with previous observations \cite{Komada_2020}. The splitting reflects extremal orbits from two equivalent pockets in nodal rings that are turned by an angle of 90 deg with respect to each other in the BZ.

Figure \ref{fig:angdep2} gives a zoom-in at low fields of the torque data from Figure \ref{fig:angdep}c. The small-frequency branch observed in the magnetisation also appears. Additionally, a second, slightly higher frequency clearly emerges for magnetic fields close to the [100] direction as visible in the raw data (for example in the 90\,deg data in Figure \ref{fig:angdep}c the oscillation with maxima at roughly 4.5\,T, 6.5\,T and 9.5\,T or in the 80\,deg data in Figure \ref{fig:angdep2} for $B > 0.6$\,T). When the frequency is larger than 5\,T, this oscillation also appears as a peak in the FFT (upward triangles in Figure \ref{fig:angdep}d). The small oscillation frequencies were extracted from the raw data as well as the FFT, where possible.
Note that we do not observe the small frequency or a signature of its quantum limit in the resistivity data or the FFTs in Figure \ref{fig:angdep}a and b.


The resulting angle dependent frequencies are shown in Figure \ref{fig:angdepfreq}(c). In addition to the circles showing frequencies obtained from torque measurements and triangles from resistivity, stars denote frequencies determined from magnetisation. 

\section{Discussion}

In the following, we compare the experimental angle dependence with the DFT predictions. 
Comparing the bandstructure obtained with different functionals allows to give a lower-bound estimate on the error bar of the DFT: We find that the bands near the Fermi energy can be shifted in energy by a few meV with respect to each other for different functionals. Because some of the Fermi surface pockets are so small this can not only result in different Fermi surface shapes but also in Lifshitz transitions, i.e., the appearance or disappearance of entire small pockets. Fermi surfaces with cross sections of the order of 1\,T in certain field directions are therefore at the limit of resolution.

In principle, all members of the TaAs family should be compensated systems and have an equal number of electrons and holes. Previous work claimed the materials to have small intrinsic doping (probably from impurities), as the best fit of the angular dependent quantum oscillations and the Hall effect corresponded consistently to a shift of the Fermi energy to the electron doped-side for both TaP and TaAs \cite{arnold_chiral_2016,arnold_negative_2016}. However, the DFT was unable to give the experimental angular dependence for some of the pockets in those works and therefore is of limited value for giving the correct electron count. A major argument, that a better compensation might actually be present in the real materials is the insensitivity of the quantum oscillation frequencies to different growth conditions. Many independent experimental quantum-oscillation studies on TaP, TaAs and NbAs show very similar quantum oscillation frequencies, although differences seem to be higher for Nb compounds compared to Ta compounds and quite strong differences were found for NbP \cite{arnold_chiral_2016,arnold_negative_2016,Naumann_2020,Laliberte_2020,Nair_2020,Komada_2020}.
Having the above points in mind, we find a relatively good agreement of the calculated and the experimental angular dependence for GGA without shifting the Fermi energy away from the compensated level. The resulting Fermi surface is shown in Figure \ref{fig:angdepfreq}(a) and (b).

For this solution, there are nested electron pockets E1 and E2 and nested hole pockets H1 and H2. Compared to TaAs, the inner nested pockets appear because of a smaller SOC. The E1 pockets are predicted to host the W1 Weyl point pairs, resulting in pockets with zero Chern number. Additionally one finds pockets H1' and E1' arising from the same bands as H1 and E1. The pockets W2 individually enclose single W2 Weyl nodes and are thus predicted to be Weyl pockets. In the angular dependence, E2, H1',W2 and E1' have frequencies in the range of the observed low quantum oscillation frequencies. In TaAs, both H1' and W2 pockets also exist and are larger. Note that the E1' pocket by symmetry should result in one more branch lying close to the lowest one in the angular dependence for all directions. This is however beyond the resolution of our calculations.

The agreement of the calculated H1 and E1 pockets with the experiment is rather good (Figure \ref{fig:angdepfreq}(c)). The experimental oscillations with small frequency of $\approx 1$\,T up to $\approx 15$\,T near $B\parallel[001]$ show an angular dependence that seems to follow some branches of the W2 pocket (or the H1' pocket that has a very similar shape). However, several of the small pockets from the calculation give branches that are qualitatively similar to the observed ones. Fermi volume changes for the W2 pocket needed to get a better agreement with the experiment are within the uncertainty of the calculations. The overall agreement of the calculation with the experiment suggests that NbAs has separate Fermi surface pockets around the W2 Weyl nodes with non-trivial topology and a Fermi energy that is only $\approx 4$\,meV above the W2 Weyl nodes. 
This distance is the smallest one reported for this class of Weyl semimetals and less than half the energy distance of the Weyl nodes to the Fermi energy in TaAs and TaP. Although it was predicted for TaAs that the chiral contribution to the conductivity should dominate the transport when the Fermi energy is less than 5 meV away from the Weyl nodes \cite{johansson_chiral_2019}, this doesn't seem to be a guarantee for the observation of a chiral anomaly in NbAs, since a clear negative longitudinal magnetoresistance from the chiral anomaly could not be identified \cite{Naumann_2020}.

It would be interesting to investigate further the quantum limit of the smallest Weyl pocket compared to the other pockets in light of the differences expected for the quantum limit magnetisation of trivial and Weyl electrons \cite{moll_magnetic_2016,Modic_2019}. However, discussing the sign of a signal in torque requires taking the different anisotropies of the Fermi surface pockets into account.

\section{Summary and Conclusion}
We investigated quantum oscillations appearing in the magnetisation, magnetic torque and resistivity in NbAs. Summarizing the above findings, the Fermi surface of NbAs consists of at least 3 pockets, one of which was not experimentally observed before. Comparing the angular dependence of the quantum oscillation frequencies with the calculated electronic structure reveals a reasonably good agreement without shifting the Fermi energy. The smallest pocket showing quantum oscillations of the order of 1\,T most probably stems from Weyl pockets around the W2 Weyl nodes which are only 4\,meV below the Fermi energy. Hence, NbAs likely hosts topological Weyl pockets.

\section{Acknowledgements}
E.H. thanks K. Behnia and B. Fauqu\'e for insightful discussions.
The authors would like to acknowledge M. Nicklas, M. Baenitz, H. Borrmann and D. Sokolov for experimental support and fruitful discussion as well as the Max-Planck society for their support of the research group "Physics of Unconventional Metals and Superconductors". E.H. and M.A.W. acknowledge support from the DFG through TRR80 (project number 107745057, project E3). M.A.W. was supported by DFG GACR project WI3320/3-1.

%

\end{document}